\documentclass[aps,prl,reprint,superscriptaddress,amsmath,amssymb]{revtex4-2}

\usepackage[version=3]{mhchem}
\usepackage{float}
\usepackage{graphicx}

\begin{document}

\title{The Microscopic Nature of Orbital Disorder in \ce{LaMnO3}}

\author{Bodoo Batnaran}
\author{Andrew L. Goodwin}
\author{Michael A. Hayward}
\author{Volker L. Deringer}
\email{volker.deringer@chem.ox.ac.uk}
\affiliation{Inorganic Chemistry Laboratory, Department of Chemistry, University of Oxford, Oxford OX1 3QR, UK}

\begin{abstract}
We present a revised atomistic picture of the order--disorder transition in the archetypal orbital-ordered perovskite material, \ce{LaMnO3}. 
Our study uses machine-learning-driven molecular-dynamics simulations which describe the temperature evolution of pair distribution functions in close agreement with experiment. 
We find the orbital-disordered phase in \ce{LaMnO3} to comprise a mixture of differing structural distortions with and without inversion symmetry, implying a mixture of different orbital arrangements. 
These distortions are highly dynamic with an estimated lifetime of $\sim 40$~fs at 1,000~K, and their fluctuations converge with the timescales of conventional thermal motion in the high-$T$ phase -- indicating that the electronic instability responsible for static Jahn--Teller distortions at low temperature instead drives phonon anharmonicity at high temperatures.
Beyond \ce{LaMnO3}, our work opens an avenue for studying a wider range of correlated materials.
\end{abstract}

\maketitle

Orbital correlations can play a key role in determining electronic and magnetic properties of materials. In many transition-metal compounds, the occupation of $d$-orbitals and their orientations control overlap, which influences electronic transport and magnetic coupling. These orbital arrangements can be  crystallographically ordered or disordered, and the balance between the two extremes can be critical to emergent phenomena such as superconductivity \cite{Bednorz1986a, Keller2008}, colossal magnetoresistance \cite{Millis1995, Millis1996, Deisenhofer2005}, charge-orbital ordering \cite{Radaelli1997}, and ferroelectricity \cite{Senn2016a}. Therefore, understanding orbital-ordered and -disordered states in crystalline solids is an important research objective. 

While the long-range ordered structural distortions induced by orbital ordering can be directly deduced from bulk diffraction studies \cite{Okazaki1961, Rodriguez1998}, it is much more difficult to describe orbital-\emph{disordered} states, both experimentally and computationally. In general, three pictures are usually considered for orbital-disordered phases. The first is that of an orbital liquid or glass, where local orbital correlations resemble those of a low-temperature ordered state, but correlated only over finite lengthscales (\emph{e.g.}, domains) \cite{Zhou1999a, Qiu2005}; the second is a correlated-disorder model, in which the disordered state supports orbital correlations not found directly in the related ordered state \cite{Ahmed2009}; and the third involves suppression of structural distortions through a displacive transition \cite{Jiang2023, Genreith2024, Nagle-Cocco2024a}. Differentiating these descriptions requires probes that are sensitive to the local distortions in the orbital-disordered phase and their corresponding dynamics.

The archetypal orbital-ordered system is \ce{LaMnO3} \cite{Zhou1999a}, which contains Mn$^{3+}$ with a $(t_{2g})^{3}(e_{g})^{1}$ high-spin configuration. Under ambient conditions, the compound adopts a perovskite structure in which each [\ce{MnO6}] octahedron is Jahn--Teller (JT) distorted to have two long, two medium, and two short Mn--O bonds. The distortions order such that two opposing long Mn--O bonds alternate orientations between two orthogonal directions, in a way that repeats along a third direction (``C-type'' ordering) \cite{Pavarini2010}. This anisotropy is evident in the orthorhombic strain of \ce{LaMnO3}. Above 750~K, the lattice parameters collapse to yield a pseudo-cubic phase in which the JT distortion is apparently quenched \cite{Rodriguez1998}. However, local-structure probes show that the asymmetry in Mn--O bond-length distributions remains after the transition, which was interpreted as there being no change in the Mn--O bond lengths with increasing $T$. The implication is that the JT distortions persist, but now in a crystallographically disordered manner which, on average, gives the observed pseudo-cubic phase \cite{Sanchez2003, Qiu2005}. 

Experiments have yielded conflicting descriptions of this order--disorder transition. Resistivity and thermoelectric measurements of the orbital-disordered phase of \ce{LaMnO3} led to an interpretation involving dynamic cooperative JT distortions, in which the C-type order fluctuates above 750~K \cite{Zhou1999a}. Additionally, a model containing nanoregions of static C-type order was proposed based on neutron pair distribution function (PDF) analysis \cite{Qiu2005}. In contrast, a three-state Potts (``3SP-type'') model, which allows the two opposing long Mn--O bonds to align along all three principal axes, was found to reproduce the behavior of \ce{LaMnO3} in resonant X-ray scattering \cite{Murakami1998, Ahmed2009}. Furthermore, reverse Monte-Carlo (RMC) fits to PDFs calculated from X-ray and neutron diffraction experiments yielded a model with the long bonds in the JT-distorted octahedra aligning perpendicularly to each other (``L-type'' ordering) \cite{Thygesen2017a}. Notably, the PDFs showed a discontinuous change at 750~K, indicating that the high-$T$ phase has a different local structure to the ordered one. However, the conclusions were not definitive as the differences between fits to high-$T$ PDF data using C-type, 3SP-type, and L-type models were not meaningful in the presence of strong thermal fluctuations \cite{Thygesen2017a}. 

The orbital order--disorder transition in \ce{LaMnO3} is similarly difficult to interpret using computations. First-principles studies indicated that the transition is multifaceted, involving a combination of octahedral rotations, antipolar distortions of the A-site cations (La$^{3+}$)  and, importantly, a compressive tetragonal shear strain on the lattice stabilizing the cooperative JT distortions in the low-$T$ antiferromagnetic phase \cite{Schmitt2020}. Comparing the computed energies of different magnetic orderings in the high-$T$ paramagnetic phase shows that antiferromagnetic (AFM) ordering yields a stronger stabilization of the JT distortion modes than ferromagnetic ordering. This stabilization results in an improved prediction of the transition temperature, implying that spin symmetry breaking, and thus dynamics, affect the transition. While these simulations have allowed the study of distortion modes and magnetic ordering in \ce{LaMnO3}, they have been limited to relatively small system sizes.

\begin{figure}[t]
 \centering
 \includegraphics[width=8.5cm]{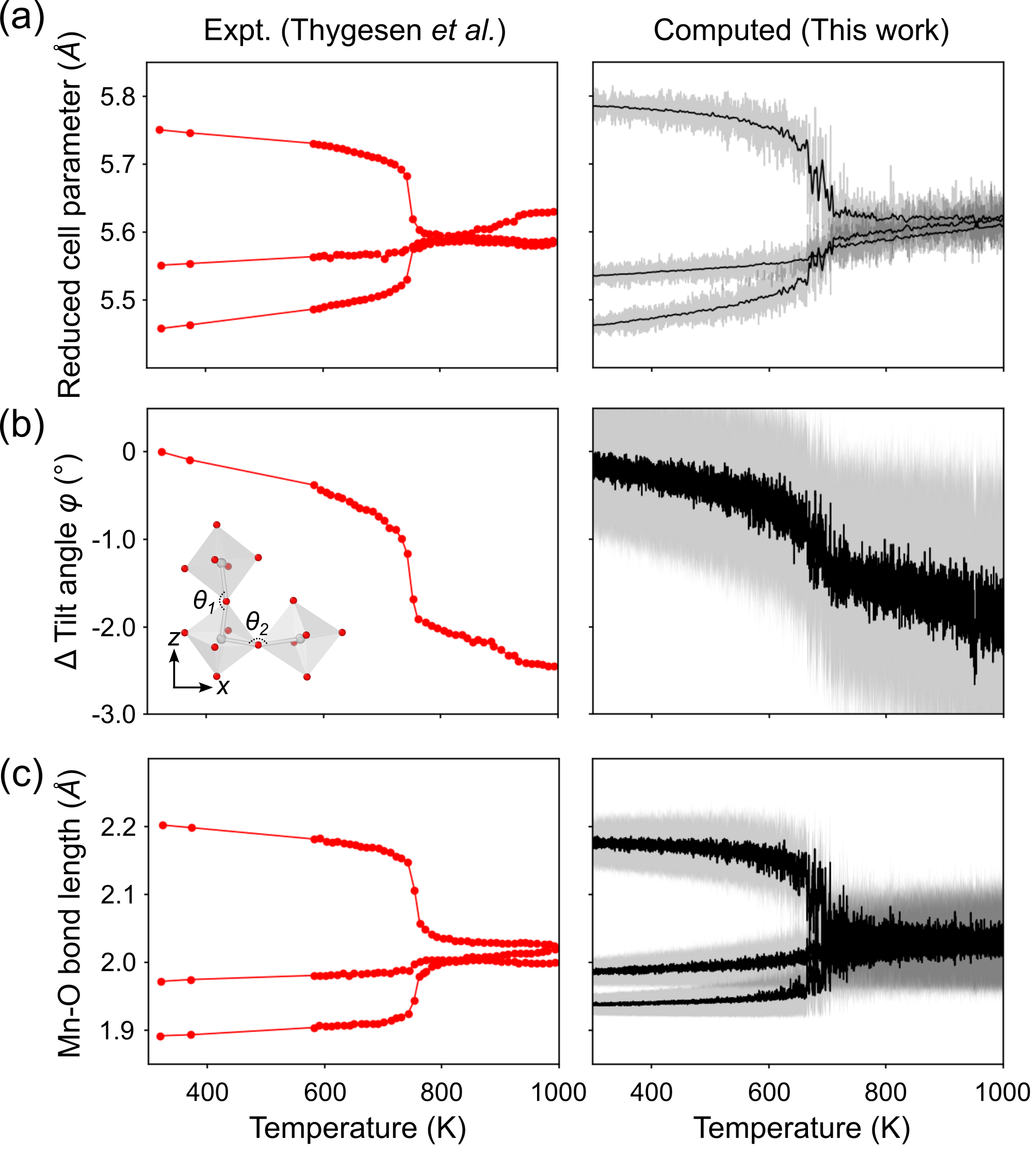}
 \caption{Change in average structure of \ce{LaMnO3} with temperature. Experimental (red) and simulated (black) data are shown. 
 (a) Reduced lattice parameters. 
 (b) Change in tilting angle, $\varphi$, of octahedra along the (111) axis. 
 (c) Mn--O bond lengths showing two long, two medium, and two short bonds and their evolution. Experimental data taken from Ref.~\citenum{Thygesen2017a}.}
 \label{fgr:props}
\end{figure}

Here, we use large-scale, machine-learning-driven molecular-dynamics (MD) simulations to re-examine the microscopic nature of orbital order and disorder in \ce{LaMnO3}. We first validate our simulations by computing the temperature evolution of average structural parameters and PDFs, comparing to experimental data. We then analyze the evolution of JT distortions with temperature, demonstrating that the high-$T$ phase of \ce{LaMnO3} actually adopts a mixture of the orbital orderings from the 3SP-type and L-type models. Finally, by computing the time-dependent evolution of the amplitudes of JT distortions, we show that these distortions become dynamic with increasing temperature and that their lifetimes shorten to the same timescale as thermal fluctuations at 1,000~K. Hence the apparent persistence of JT distortions in orbital-disordered \ce{LaMnO3} is really a signature of JT-driven bond anharmonicity.

\begin{figure}[t]
 \centering
 \includegraphics[width=8.5cm]{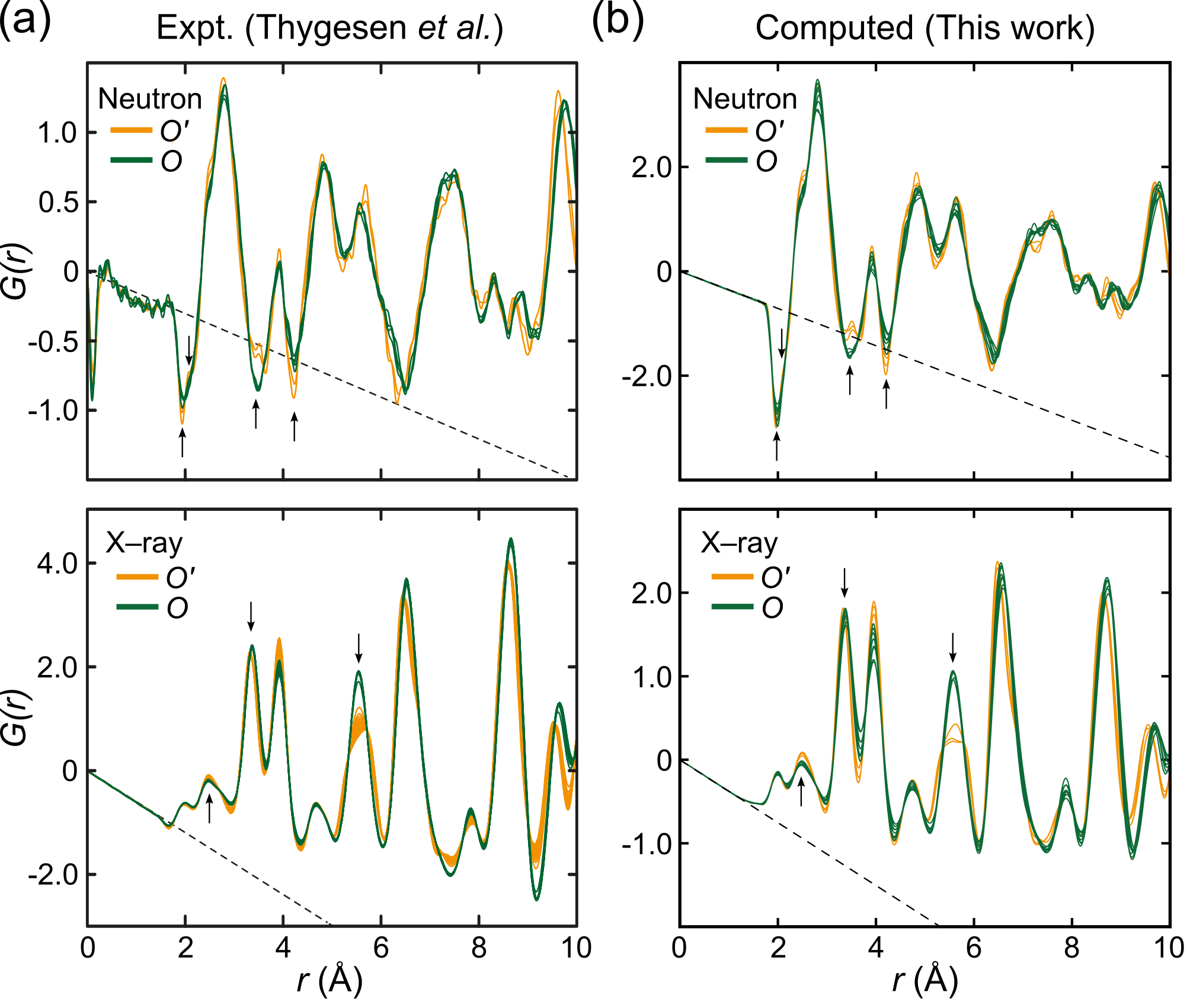}
 \caption{Pair distribution functions for \ce{LaMnO3} and their evolution with temperature. Experimental (panel a) and computed (panel b) neutron (top) and X-ray (bottom) PDFs for the temperature range 500 to 1000~K are shown. Baselines are shown using dotted lines; features of particular interest are marked with arrows. As in Ref.~\citenum{Thygesen2017a}, \textit{O'} (orange) denotes the orbital-ordered low-$T$ and \textit{O} (green) denotes the orbital-disordered high-$T$ phase. Panel (a) adapted from Ref.~\citenum{Thygesen2017a}.}
 \label{fgr:pdf}
\end{figure}

Our MD simulations are driven by a machine-learned interatomic potential (MLIP). Recent advances in on-the-fly-accelerated first-principles MD \cite{Li2015a, Jinnouchi2019, Stenczel2023} provide a route for the efficient generation of training data for MLIPs \cite{Liang2023, El-Machachi2024}. We here parameterized an MLIP model using the NequIP architecture \cite{Batzner2022a, footnote-nequip} together with an existing dataset of 1,384 \ce{LaMnO3} structures from Ref.~\citenum{Jansen2023}, generated initially from a 100 $\rightarrow$ 1,100~K heating run of a  $2 \times 2 \times 2$ cubic unit cell with AFM ordering \cite{Jansen2023}. For MD simulations, the isothermal--isobaric (NPT) ensemble was used with a Nos\'e{}--Hoover thermostat and barostat as implemented in LAMMPS \cite{Thompson2022a}.

To validate the MLIP model, we computed key structural parameters and compared their change with temperature to the experimental data of Ref.~\citenum{Thygesen2017a} (Fig.~\ref{fgr:props}). A $6 \times 6 \times 6$ expansion of the cubic unit cell was used for a heating simulation from 200 to 1,200~K at a rate of 1~K/ps at constant atmospheric pressure. Reduced cell parameters were calculated using equations outlined in Ref.~\citenum{Jansen2023}, and the tilting angle, $\varphi$, along the (111) axis was calculated using the O--O--O angles $\theta_1$ and $\theta_2$ according to the methods in Refs.~\citenum{Okeeffe1977} and \citenum{Parsons2009a}. The average structural parameters were obtained from a trajectory with a time averaging of 0.1 ps. For the lattice parameters, a moving average was calculated and overlaid on the time-averaged values. The tilting angles and bond lengths were averaged over all 256 octahedra within the supercell. Overall, Fig.~\ref{fgr:props} shows excellent agreement between simulated and experimental data: in particular, the orthorhombic/pseudocubic transition associated with orbital order/disorder is replicated. The temperature at which this transition occurs is underestimated by $\sim 60$~K, which may be due to the underlying DFT level \cite{Schmitt2020, Jansen2023}. With increasing $T$, the lattice parameters collapse to similar values accompanied by a decrease in shear strain \cite{LaMnO3-GitHub-Repo}, the octahedral tilting decreases, and the JT distortions appear to be suppressed. At $\sim 1,200$~K, the tilting pattern changes to $a^{-}a^{-}a^{-}$, consistent with a transition to the rhombohedral phase as observed experimentally \cite{LaMnO3-GitHub-Repo}. 

Next, we calculated averaged PDFs from structures sampled from 10~ps MD runs. Neutron and X-ray PDFs were predicted using DebyeCalculator \cite{Johansen2024} using reciprocal-space ranges, real-space resolutions, and damping parameters from Ref.~\citenum{Thygesen2017a}. Figure~\ref{fgr:pdf} shows that the calculated PDFs agree remarkably with experimental data from Ref.~\citenum{Thygesen2017a}, indicating that the simulation captures the change in local structure within the temperature range. Notably, it replicates the change in peaks corresponding to the Mn--O bonds, from two split peaks to a broad single peak at $\sim 2$~\AA. While the changes in lattice parameters, volume, and Mn--O bond lengths have been previously simulated \cite{Sartbaeva2006, Sartbaeva2007, Ahmed2009, Schmitt2020}, this is the first time that the experimental PDFs for \ce{LaMnO3} -- and their evolution across the order/disorder transition -- have been replicated from first principles.

\begin{figure}[t]
 \centering
 \includegraphics[width=8.5cm]{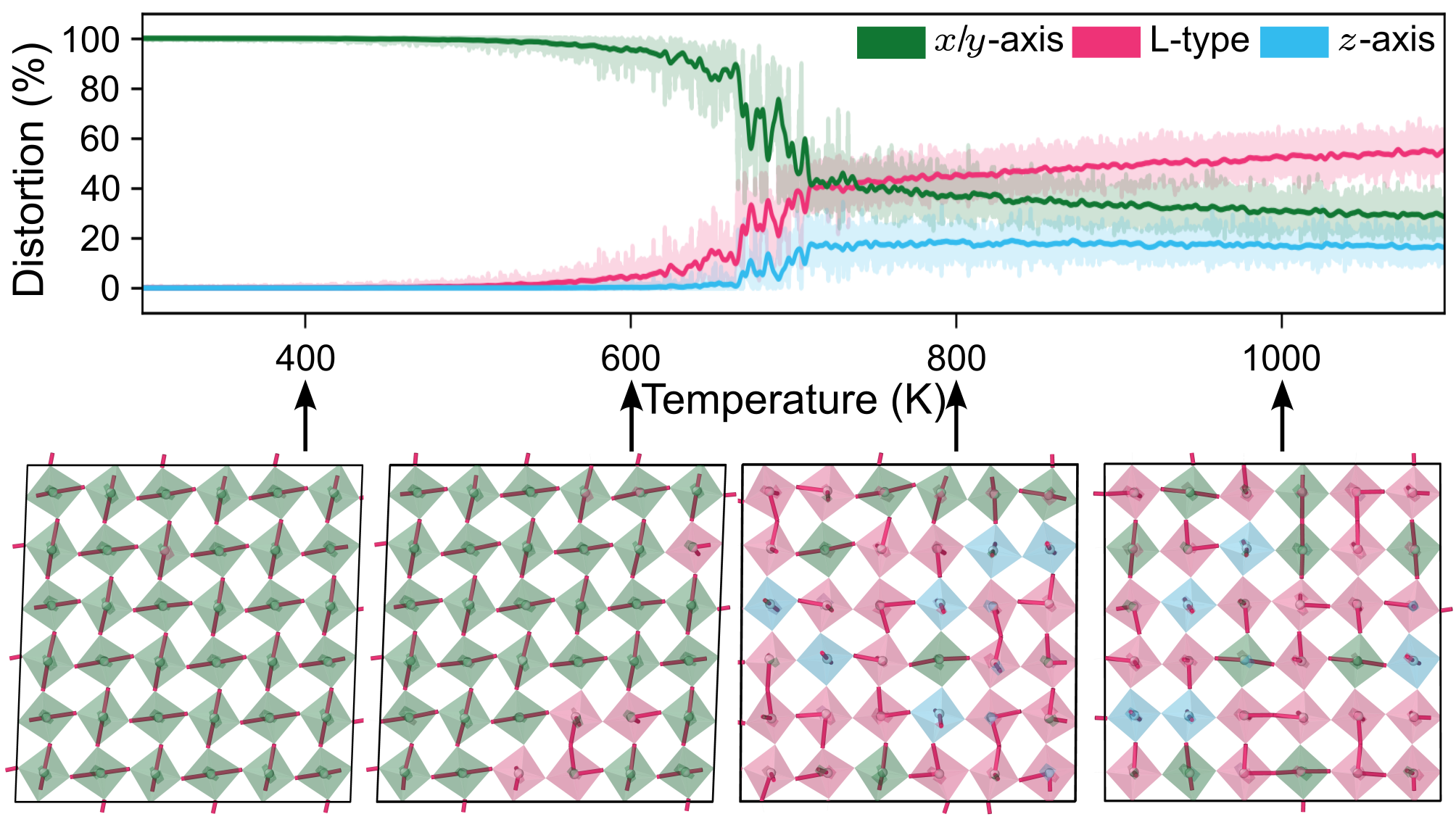}
 \caption{Evolution of long Mn--O bond orientations with temperature. We plot the percentage of parallel long bonds aligning with the $x$- or $y$-axis (green) and $z$-axis (cyan) as well as L-type (magenta). Slices in the $xy$-plane highlighting the two longest bonds are shown at relevant temperatures \cite{Stukowski2009}.}
 \label{fgr:model}
\end{figure}

Overall, Figs.~\ref{fgr:props} and \ref{fgr:pdf} show that the temperature dependence of both average and local structure in \ce{LaMnO3} is well described by our simulations. This observation justifies now using the atomistic configurations to explore the nature of orbital order and disorder in \ce{LaMnO3}. Figure~\ref{fgr:model} shows the orientations of the respective two longest Mn--O bonds in each \ce{[MnO6]} octahedron with temperature, as well as snapshots of slices in the $xy$-plane. The MD trajectory has been time-averaged with a window of 0.1~ps to remove thermal fluctuations. We found that larger time-averaging windows resulted in the bond lengths averaging out in the high-$T$ phase due to the constant fluctuation of long-bond orientations \cite{LaMnO3-GitHub-Repo}. As temperature rises, the system undergoes a drastic change, losing the C-type order and adopting a crystallographically disordered configuration. Unexpectedly, the high-$T$ phase consists of $\sim 50\%$ of L-type distortions and the remaining distortions having opposing long bonds aligning along all three axes in equal proportions. This result is consistent with the meaningful change in local structure observed in experimental PDFs \cite{Thygesen2017a}, but is a markedly different picture than previously considered.

\begin{figure}[t]
 \centering
 \includegraphics[width=8.5cm]{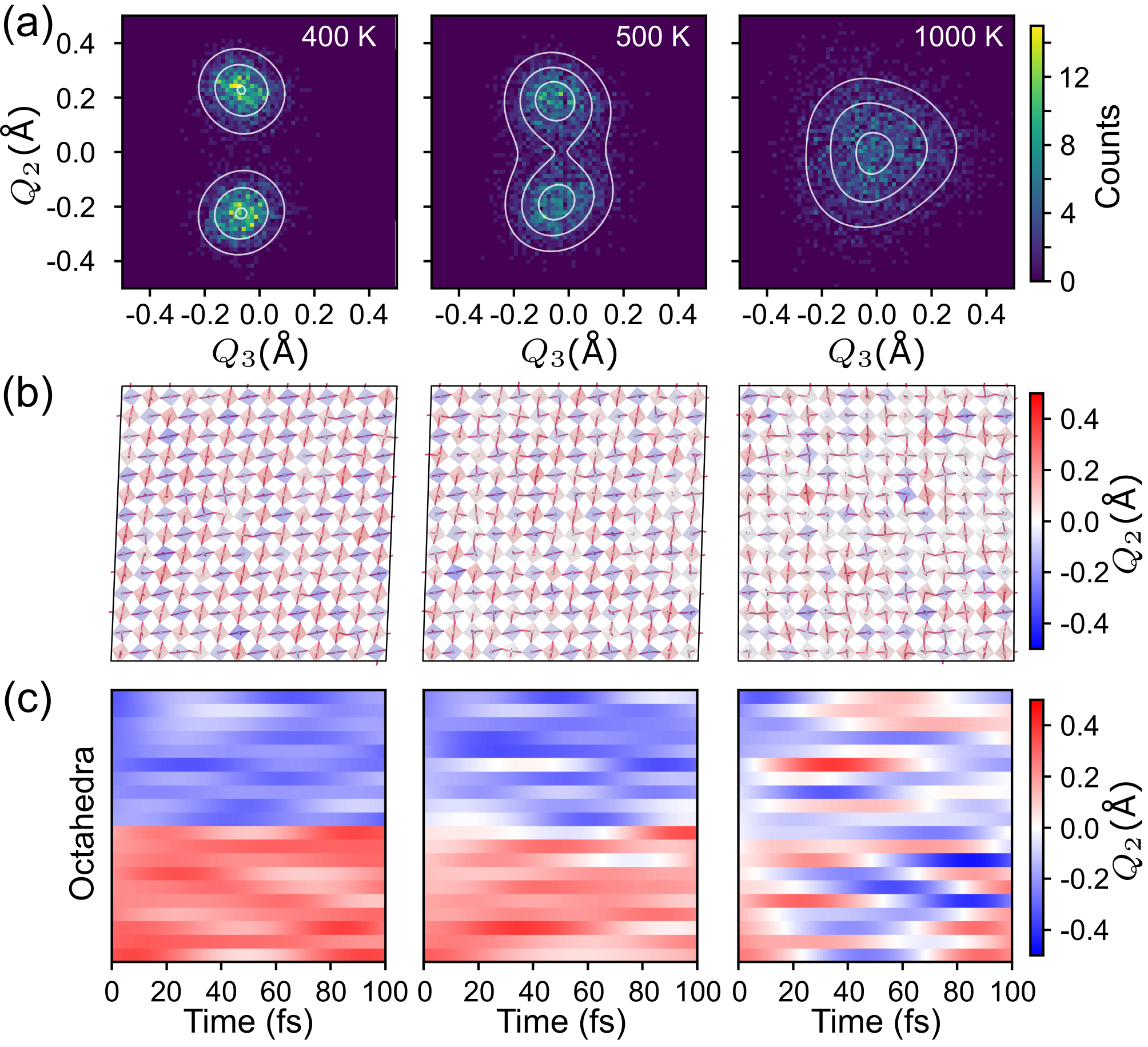}
 \caption{Evolution of octahedral distortions in a 13,720-atom supercell with temperature and time. 
 (a) Heatmaps with contour lines of the octahedral distortion mode components, $Q_2$ and $Q_3$, at 400~K, 500~K and 1,000~K. 
 (b) Slices along the $xy$-plane of the supercells \cite{Stukowski2009}. The amplitude of the $Q_2$ component for each octahedra and Mn--O bonds longer than 2.05~\AA\ are colored. 
 (c) Change in the amplitude of the $Q_2$ component with simulation time for 20 randomly selected octahedra, ordered by the initial amplitude, at each temperature.}
 \label{fgr:grid}
\end{figure}

To draw a comparison to other materials, a clear change in local structure is also apparent from PDFs for orbital order transitions in the layered nickelates, \ce{LiNiO2} \cite{Genreith2024} and \ce{NaNiO2} \cite{Nagle-Cocco2024a}: the Ni--O peaks broaden and merge with increasing temperature, similar to the Mn--O peaks in \ce{LaMnO3}. However, the change in the Ni--O peaks was interpreted to be due to the JT distortions disappearing, in contrast to the canonical understanding of the behavior of \ce{LaMnO3}. We can investigate this difference in interpretation by calculating the magnitude of the JT distortions of each octahedron through Van Vleck vibration modes, using VanVleckCalculator \cite{Nagle-Cocco2024b}. We focus on the $E_g$ distortion space which is spanned by a planar rhombic $Q_2$ component (compression of two bonds and stretching of two bonds) and a tetragonal $Q_3$ component (compression of four bonds and stretching of two bonds). Figure \ref{fgr:grid}a shows the evolution of the distributions of the octahedral distortions from snapshots of a  $14 \times 14 \times 14$ supercell heated for 10~ps at each temperature. At 400~K, there are two high-intensity regions showing that there are two major orientations of the JT distortions, as expected for C-type ordering. At 500~K, these regions are broader and closer together in distortion space. When slices of the 500~K configurations are visualized with color-coding based on the amplitude of the $Q_2$ component (Fig.~\ref{fgr:grid}b), we see mostly C-type ordering, but a small proportion of the octahedra have near-zero $Q_2$ amplitudes. At 1,000~K, the heatmap in Fig.~\ref{fgr:grid}a shows a clear centering around $(0, 0)$, and the majority of the octahedra have reduced $Q_2$- or $Q_3$-amplitudes ($\sim$ 45\% within $\pm$0.05~\AA). Slices of the 1,000~K configurations show essentially complete melting of the original C-type order. 

The unimodal distribution of octahedral distortion types we observe is inconsistent with previous interpretations based on the 3SP model \cite{Sartbaeva2007}, for which the ($Q_2, Q_3$) heatmap would show three maxima at high temperature, corresponding to JT distortions aligned along the three principal axes. Hence, our results indicate that the high-temperature phase of \ce{LaMnO3} involves suppression of the conventional JT distortion of \ce{[MnO6]} octahedra on long timescales. The broadening and merging of the Mn--O peaks in the PDFs is probably a signature of this local-structure transition. Indeed, this interpretation has much in common with that of JT suppression in \ce{LiNiO2} \cite{Genreith2024} and \ce{NaNiO2} \cite{Nagle-Cocco2024a}.

Next, we ask whether the octahedral distortions in the high-$T$ phase are static, as had been inferred from experimental studies \cite{Qiu2005, Ahmed2006}. We qualitatively analyzed JT mode lifetimes by tracking the amplitude of the $Q_2$ component of each octahedron over time. Figure~\ref{fgr:grid}c compares the lifetimes at different temperatures for 20 randomly selected octahedra. Each horizontal line in the plot represents the evolution of the $Q_2$ component of one octahedron in the last 100~fs from 10~ps runs at constant $T$. While essentially static at 400~K, the lifetimes shorten and the distortions start to fluctuate between different directions at 500~K (red $\leftrightarrow$ blue in Fig.~\ref{fgr:grid}c). At 1,000~K, the distortions are constantly swapping orientations with a lifetime of $\sim 40$~fs \cite{LaMnO3-GitHub-Repo}, showing that distortions in the high-$T$ phase are highly dynamic. This behavior is reminiscent of JT-driven fluctuations in paraelectric \ce{BaTiO3}, in which asymmetric broadening of the Ti--O PDF peaks at high temperatures is associated with anharmonicity of Ti-off-centering dynamics \cite{Senn2016a, Pasciak2018}. Similar interplay of JT instabilities and strongly anharmonic lattice dynamics is found in the thermoelectrics PbTe and SnSe  \cite{Bozin2010, Delaire2011, Sangiorgio2018, Jiang2023}. For PbTe, single-crystal inelastic neutron scattering showed that the electronic instability causes phonon anharmonicity resulting in dynamic dipoles that give asymmetric PDF peaks with increasing temperature \cite{Jensen2012}. The dynamic nature of the octahedral distortions in \ce{LaMnO3} might be confirmed experimentally by the same approach. For comparison with future studies, simulated single-crystal diffuse scattering patterns from the configurations with 13,720 atoms are provided online \cite{LaMnO3-GitHub-Repo}.

Our key result has been to show that the conventional JT distortion found in octahedral Mn$^{3+}$ perovskites is actually quenched in the orbital-disordered phase of \ce{LaMnO3}. The distortions that are present at high temperatures are more varied and fluctuate on a timescale that is similar to that of typical thermal motion. Hence the orbital arrangements in \ce{LaMnO3} are not well described by a conventional order/disorder picture, but instead involve a transition from static correlated JT to dynamic JT-driven anharmonicity. Because CMR emerges from this ``orbital-disorder'' phase on doping, our analysis suggests that MLIP-driven MD simulations may offer new insight into the microscopic nature of the CMR phenomenon. 

B.B. acknowledges discussions with C.\ Ben Mahmoud, L.A.M.\  Rosset, R.\  Wernert, and Y.\  Zhou, and funding from the EPSRC Centre for Doctoral Training in Inorganic Chemistry for Future Manufacturing (OxICFM), EP/S023828/1. A.L.G. acknowledges discussions with M.S.\ Senn, D.A.\ Keen, and L.A.V.\ Nagle-Cocco. The authors acknowledge the use of the University of Oxford Advanced Research Computing facility (http://dx.doi.org/10.5281/zenodo.22558). 
Data supporting this work are openly available at Ref.~\citenum{LaMnO3-GitHub-Repo}.

\end{document}